\documentclass[aps,prb,showpacs,reprint,amsmath,amsfonts,amssymb,superscriptaddress] {revtex4-1}
\usepackage{graphicx}

\begin{document}

\title{Suppression of dense Kondo state in CeB$_{6}$ under pressure}

\author{N.~Foroozani}
\affiliation {Department of Physics, Washington University, St. Louis, MO 
63130, USA}

\author{J.~Lim}
\affiliation {Department of Physics, Washington University, St. Louis, MO 
63130, USA}

\author{G.~Fabbris}
\affiliation {Department of Physics, Washington University, St. Louis, MO 
63130, USA}
\affiliation {Advanced Photon Source, Argonne National Laboratory, Argonne, IL 
60439, USA}

\author{P.~F.~S.~Rosa}
\affiliation { Department of Physics and Astronomy, University of California,
Irvine, CA 92697, USA}

\author{Z.~Fisk}
\affiliation { Department of Physics and Astronomy, University of California,
Irvine, CA 92697, USA}

\author{J.~S.~Schilling}
\email[Corresponding author: ]{jss@wuphys.wustl.edu}
\affiliation {Department of Physics, Washington University, St. Louis, MO 
63130, USA}

\date{\today}

\begin{abstract}

To investigate whether the dense Kondo compound CeB$_{6}$ might evolve into a
topological insulator under sufficient pressure, four-point electrical
resistivity measurements have been carried out over the temperature range 1.3
K to 295 K in a diamond anvil cell to 122 GPa. The temperature $T_{\max}$ of
the resistivity maximum initially increases slowly with pressure but
disappears between 12 and 20 GPa. The marked changes observed under pressure
suggest that a valence and/or structural transition may have occurred.
Synchrotron x-ray diffraction measurements, however, fail to detect any change
in crystal structure to 85 GPa. Although a transition into an insulating phase
is not observed, this dense Kondo system is completely suppressed at 43 GPa,
leaving behind what appears to be a conventional Fermi liquid metal.

\end{abstract}

\maketitle

\section{Introduction}

Many compounds and alloys containing Ce, including Ce metal, show highly
anomalous magnetic properties due to the fact that the 4$f$ level in trivalent
Ce often lies near a magnetic instability. This led to the observation of a
classic Kondo effect in the electrical resistivity measurements of Winzer
\cite{winzer1} on the dilute magnetic alloy (La$_{0.994}$Ce$_{0.006}$)B$_{6}$
to 50 mK; in these studies the Kondo temperature is estimated to lie near
$T_{\text{K}}\approx$ 1 K. At higher Ce concentrations the competition between
the Kondo effect and Ce-Ce exchange interactions increases, leading finally in
CeB$_{6}$ to an exotic phase diagram with a prominent resistivity maximum near
4 K and two kinds of ordered phases: \ antiferroquadrupolar ordering below
3.2~K (phase II) and antiferromagnetic ordering below 2.3~K (phase III)
\cite{2highp,3highp}. This compound has attracted a great deal of attention
following its identification as a dense Kondo system \cite{fujita1}. CeB$_{6}$
is metallic at ambient pressure with the simple cubic (\textit{Pm3m})
structure \cite{leger1}. The ground state of the Ce$^{3+}$ ion is a
$\Gamma_{8}$ quartet split by 30 K into two doublets and separated by 540~K
from the excited $\Gamma_{7}$ doublet \cite{1kobayashi}.

SmB$_{6}$ is a prime candidate for a topological Kondo insulator categorized
as a heavy fermion semiconductor \cite{NA1,NA2,NA3,NA4} with strong
electron-electron correlations in which the localized 4$f$ electrons give rise
to novel ground states \cite{nature1,nature2}. SmB$_{6}$ and CeB$_{6}$ both
have the $J=5/2$ Hund's rule configuration \cite{1kobayashi,nyhus1}.\emph{\ }
Upon cooling below 10 K, the electrical resistivity of SmB$_{6}$ rises by
orders of magnitude as it enters into an anomalous insulating state, only to
saturate near 7 m$\Omega$cm \cite{neupane}. This behavior agrees with that of
a Kondo insulator where at high temperatures highly correlated electron
behavior is observed, but at the lowest temperatures an insulating state
emerges as a bulk band gap opens up through the hybridization of localized
4$f$ states with 5$d$ conduction electrons \cite{NA1,NA2,NA3,NA4}. Since the
primary effect of high pressure on a dense Kondo system is to enhance this
hybridization \cite{NF1}, it is conceivable that sufficient pressure might
succeed in transporting CeB$_{6}$ into the topological Kondo insulator regime.

To date there have been relatively few studies of the transport and magnetic
properties of CeB$_{6}$ at high pressures \cite{NF1,9G,NF2}. Kobayashi
\textit{et al.} \cite{NF1} report that the prominent resistivity maximum in
CeB$_{6}$ shifts from 3.6~K to $\sim$7~K at 13 GPa pressure, but there is no
evidence for a transition to a Kondo insulating state as in SmB$_{6}$
\cite{neupane}. There is also no evidence for a structural phase transition in
CeB$_{6}$ at ambient temperature to 20 GPa \cite{8G}.

In the present study we extend the previous resistivity experiments to
significantly higher pressures (122 GPa). The resistivity maximum is found to
initially shift to somewhat higher temperatures with pressure, but rapidly
decrease in magnitude, finally disappearing completely at 20 GPa. The present
experiments fail to find any evidence that CeB$_{6}$ transforms into a Kondo
insulating state to 122 GPa in the temperature range 1.3 - 295 K. However,
pressures of 43 GPa and above are sufficient to completely suppress the
anomalies associated with CeB$_{6}$'s dense Kondo state. These dramatic
changes under pressure are not due to structural phase transitions, as
evidenced by x-ray diffraction studies at 15 K to 85 GPa. However, we cannot
exclude the possibility that a pressure-induced change in Ce's valence
(Ce$^{+3}$ to Ce$^{+4}$) may have occurred.

\section{Experimental Methods}

Single crystals of CeB$_{6}$ were grown in Al flux by slow cooling from
1450$^{\circ}$C. The crystals were removed from the Al flux by leaching in
NaOH solution \cite{canfield}. The lattice parameter of the resulting simple
cubic crystals was measured to be 4.132(4)~\AA .

High pressure dc electrical resistivity measurements were carried out in a
diamond-anvil cell (DAC) using two opposing 1/6-carat, type-Ia diamond anvils
with 0.35 mm culets beveled at 7$^{\circ}$ to 0.18 mm central flats. A Re
gasket (6-7 mm diameter, 250 $\mu$m thick) was preindented to 30 $\mu$m and
insulated using a 4:1 c-BN-epoxy mixture which also served as a
quasi-hydrostatic pressure medium. The pressure was determined \textit{in situ
}by placing two small ruby spheres \cite{24N} in the sample space. Four-point
resistivity was measured using four leads cut from a thin Pt foil (see
Fig.~1); two extra leads were connected to the Re gasket to detect any
electrical shorts. A single crystal sample (dimensions 40$\times$40$\times$5
$\mu$m$^{3}$) was placed on top of the Pt leads and electrical contact was
made by pressing the sample into the leads with the opposing anvil. During the
course of the experiment the sample was plastically deformed by the
quasi-hydrostatic pressure. To keep the power dissipated in the sample below 1
$\mu$W, an excitation current of $\sim1$ mA was used. A reduction of the
current to 0.1 mA at the lowest temperature (1.3~K) caused no measurable
change in the sample resistance.

 \begin{figure}[t]
\includegraphics[width = 8.5 cm]{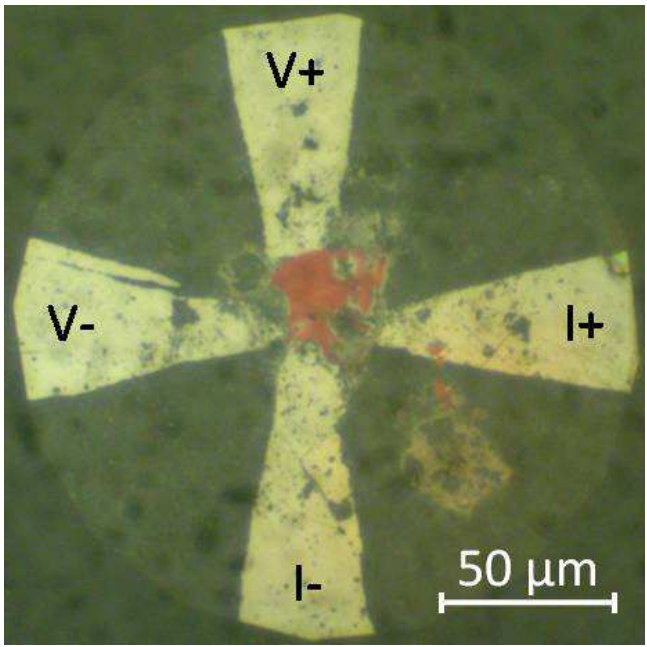} \caption{\label{fig1}(color online) Image of red CeB$_{6}$ sample
(40$\times$40$\times$5 $\mu$m$^{3}$) resting on four Pt leads (4 $\mu$m thick)
on insulated Re gasket.}
\end{figure}

A He-gas driven membrane allowed changes in pressure at any temperature above
3~K. The pressure cell was placed in a continuous-flow cryostat (Oxford
Instruments) and submerged in pumped liquid He to reach temperatures as low as
1.3~K. The pressure was determined either at ambient temperature (295~K) or at
low temperature (5-10~K) with a resolution of $\pm$ 0.2 GPa using the revised
ruby pressure scale of Chijioke \textit{et al}. \cite{25N}. Raman spectroscopy
\cite{18N} on the diamond vibron was also used to determine the pressure in
the upper pressure range where the ruby fluorescence became very weak. Further
details of the high pressure resistivity techniques are given elsewhere
\cite{16I}.

High-pressure low-temperature powder x-ray diffraction (XRD) experiments were
carried out at the HPCAT (16-BM-D) beamline of the Advanced Photon Source,
Argonne National Laboratory. A symmetric DAC (Princeton shops) was prepared
using standard diamond anvils with 300 $\mu$m diameter culets beveled to 180
$\mu$m. The anvils were glued onto WC and B$_{4}$C seats, the latter allowing
the extension of the 2$\theta$ range to $\sim$25$^{\circ}$. A Re gasket was
pre-indented to 25 $\mu$m and a hole 90 $\mu$m in diameter laser drilled
through the center of the indentation to serve as sample chamber.

Powdered CeB$_{6}$ was placed into the sample chamber together with Au powder
as pressure marker and a ruby sphere. He gas was then loaded using the
GSECARS/COMPRES system \cite{G1} to $\sim$8.6 GPa as determined by ruby
fluorescence \cite{25N}. During the high-pressure x-ray experiment the
\textit{in situ} pressure was determined from the known equation of state of
Au \cite{G3}.

The pressure cell was cooled using a He flow cryostat. Isothermal measurements
were performed at $\sim$15~K and the pressure was increased at low temperature
using a gear box. Photons of 29.2 keV energy compressed the reciprocal space,
thus allowing a significant number of Bragg reflections within the limited
2$\theta$ range. Diffraction was detected in angular dispersive mode using an
image plate (MAR3450) with 100$\times$100 $\mu$m$^{2}$ pixel size located
324.37 mm from the sample. The beam was focused to $\sim$15$\times$5 $\mu
$m$^{2}$ by a pair of Kirkpatrick-Baez mirrors. The 2D diffraction patterns
were converted to intensity versus 2$\theta$ using Fit2D software \cite{G4}.
CeB$_{6}$ and Au diffraction patterns were concomitantly fit using Le Bail and
Rietveld methods implemented with GSAS/EXPGUI software {\cite{G5,G6}.}

\section{Results of Experiment}

\subsection{Electrical resistivity measurements}

In Fig.~2 is shown resistance versus temperature for the high-pressure data
obtained on CeB$_{6}$ over the temperature range 1.3~K to 295~K. The order of
measurement is: \ 0.5, 7.7, 12, 20, \ 30, 43, 122 GPa. Above 10~K all data
shown were taken with increasing temperature due to the relatively slow rate
of warming to 295~K (12 hours). Pressure was always applied at 295 K. All
values of pressure given in Fig.~2 were measured at 295~K before cooling down;
the values of the pressure after cooling to 5-10~K and after warming back up
to 295~K, all at the same gas pressure in the membrane, are given in
parentheses in the caption to Fig.~2, if they were measured. The highest
pressure reached, 122 GPa, was determined from Raman scattering off the
diamond vibron.

 \begin{figure}[t]
\includegraphics[width = 8.5 cm]{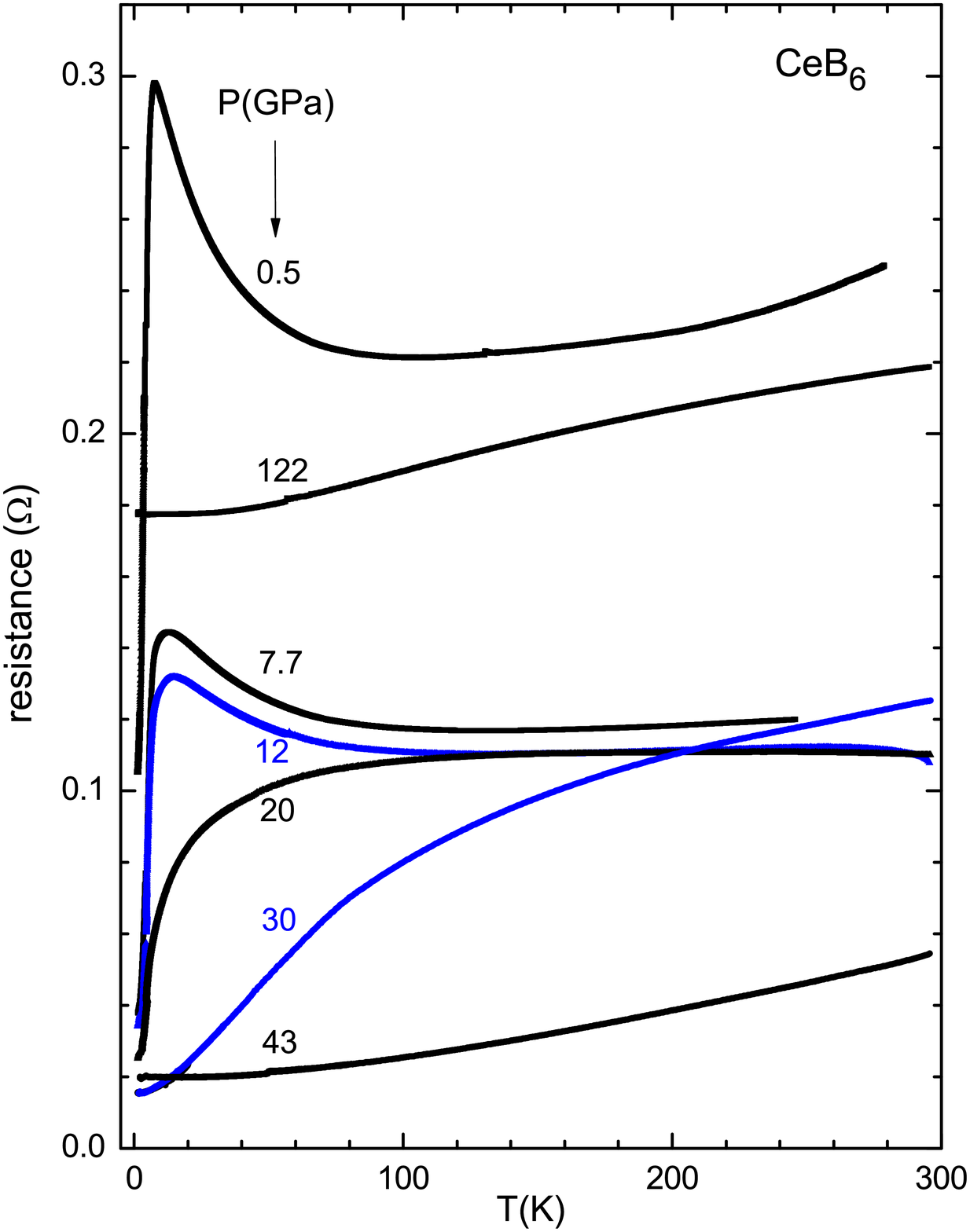} \caption{\label{fig2}(color online) Electrical resistance of CeB$_{6} $
versus temperature at pressures in GPa measured at 295~K before cooling.
Numbers in parentheses after each pressure give pressures measured at 5-10~K
after cooling (left) and at 295~K after warming back to 295~K (right):
\ 0.5(1.2,0.8), 7.7(--,8.5), 12(--,14), 20(24.5,--), 30(35,31), 43(52,45),
122(-,129). A dash "-" indicates the pressure was not measured.}
\end{figure}

At 0.5 GPa the temperature-dependent resistance $R(T)$ displays a resistivity
minimum near 110~K followed at lower temperatures by a peak near $T_{\max
}\approx6$~K, a somewhat higher temperature than $T_{\max}\approx$ 3.5 K
reported in earlier measurements on CeB$_{6}$ in the 0-1 GPa pressure range
\cite{Slu,mori1,physica B}. Kobayashi \textit{et al.} \cite{physica B} find
that $T_{\max}$ shifts to higher temperatures with pressure, reaching 7 K at
13 GPa.\ From the data in Fig.~2 one also sees that $T_{\max}$ increases with
pressure, the resistivity peak itself diminishing rapidly in size and finally
disappearing between 12 and 20 GPa. For pressures up to and including 30 GPa,
the resistance below 100~K decreases with pressure; for temperatures above
20~K the decrease in resistance between 30 and 43 GPa is particularly large.
This decrease likely reflects the intrinsic pressure dependence of the
CeB$_{6}$ sample since any plastic deformation of the sample by the solid
pressure medium would be expected to \textit{enhance} its resistance.

As the pressure was increased above 43 GPa, however, the resistance at ambient
temperature began to shift upwards and change with time at a given pressure,
indicating relaxation behavior in the pressure cell. The two ruby spheres were
also seen to move away from the sample, preventing a reliable determination of
the sample pressure. At 122 GPa the pressure was determined from the diamond
anvil vibron using Raman spectroscopy. In Fig. 2 it is seen that the entire
resistance curve $R(T)$ at 122 GPa has risen to a significantly higher value,
\textit{nine times} that at 43 GPa. The reason for the large increase in
resistance is not clear, but could be the result of a structural phase
transition with a mixed phase region, leading to enhanced defect scattering.
Alternatively, the outward flow of the pressure cell could generate a large
number of lattice defects in the sample from plastic deformation by the solid
pressure medium. In any case, to 122 GPa no sharp upturn in the electrical
resistivity of CeB$_{6}$ below 10 K was observed that might have signaled a
transition into a topological insulating state, as suggested for SmB$_{6}$
\cite{neupane}.

Following the measurement at 122 GPa, the pressure was released completely.
The $R(T)$ dependence was similar to that at 122 GPa but shifted to even
higher resistance values. In addition, the characteristic resistivity maximum
seen near 6 K at 0.5 GPa was \textit{not} recovered at ambient pressure. This
finding would be consistent with a structural phase transition at extreme
pressures that remained metastable after releasing to ambient pressure.

The dramatic changes observed in the temperature-dependent resistivity of
CeB$_{6}$ under pressure from 0.5 to 43 GPa suggest that a change in valence
and/or a structural transition may have taken place. Previous diffraction
studies by Leger \textit{et al.} \cite{8G} at ambient temperature revealed no
phase transition to 20 GPa. High pressure x-ray diffraction and x-ray
absorption near-edge structure (XANES) studies to pressures of at least 50 GPa
would be helpful to clarify the situation. We next discuss the results of our
recent x-ray diffraction measurements on CeB$_{6}$ to pressures exceeding 50 GPa.

\subsection{X-ray diffraction measurements}

The x-ray diffraction pattern of CeB$_{6}$ at 15 K for selected hydrostatic
pressures with He pressure medium is shown in Fig.~3. Neither the emergence of
new Bragg peaks, nor the splitting of existing peaks, was observed to the
highest pressure (85 GPa), indicating that the structure of CeB$_{6}$ remains
simple cubic throughout this range.

 \begin{figure}[t]
\includegraphics[width = 8.1 cm]{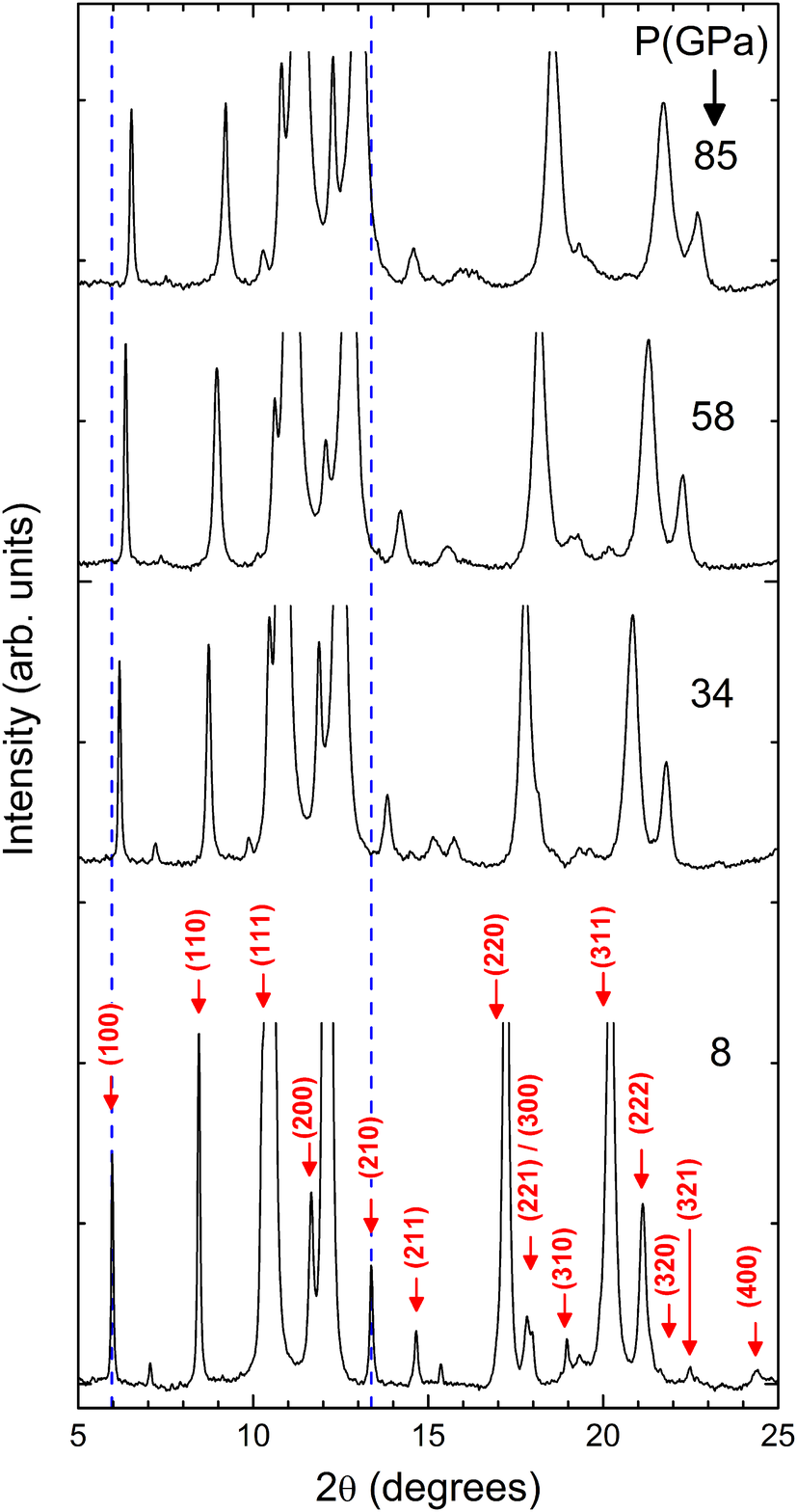} \caption{\label{fig3}(color online) CeB$_{6}$ diffraction pattern at 15
K and selected pressures. Large, truncated peaks are from Au marker. Red
arrows and indices give positions of CeB$_{6}$ peaks; all other peaks can be
assigned to Au, ruby or Re.}
\end{figure}

The pressure dependence of the unit cell volume is well described by a 3rd
order Birch-Murnaghan equation of state (see Fig.~4):
\begin{equation}
\begin{split}
P(V)=\frac{3B_{o}}{2} \left[ \left( \frac{V_{o}}{V} \right) ^{7/3} - \left( \frac{V_{o}}{V} \right)  ^{5/3} \right] \times \\
\left\{ 1+\frac{3}{4} \left( B_{o}^{\prime}-4 \right) \left[ \left( \frac{V_{o}}{V} \right) ^{2/3} - 1 \right] \right\},
\end{split}
\end{equation}
where $P$ and $V$ are the measured pressure and volume, respectively. The fit
to the data using Eq. 1 yields the ambient pressure volume $V_{o}$, bulk
modulus $B_{o}$, and pressure derivative of the bulk modulus $B_{o}^{\prime}
$. The fit parameters are summarized in Table I together with those reported
in the literature. The first fit includes only the present $V(P)$ data
measured from 8 to 85 GPa. The resulting black fit curve is seen in Fig. 4 to
lie slightly below the ambient pressure volume $V_{o}$ from Sirota \textit{et
al}. \cite{7G}.\ In Table I it can be seen that the value of $V_{o}$ lies
below the literature values, whereas the bulk modulus $B_{o}$ lies above these
values. Much better agreement is achieved if Eq. 1 is fit to the lowest 5 high
pressure points to 20 GPa in Fig. 4, including the ambient pressure point from
Sirota \textit{et al}. \cite{7G}. This excellent agreement is not surprising
since the literature studies were all carried out at pressures at or below 20 GPa.

Although no structural phase transition was detected to 85 GPa pressure, one
cannot exclude the possibility of a change in structure in the region 85 - 122
GPa that led to the strong increase in $R(T)$ at 122 GPa.

\section{Discussion}

Pressures exceeding 43 GPa cause drastic changes in the temperature-dependent
resistance $R(T)$ of CeB$_{6}$ that indicate a complete suppression of its
dense Kondo state. The prominent resistivity peak near ambient pressure
disappears between 12 and 20 GPa and $R(T)$ is strongly suppressed over nearly
the entire measured temperature range as the pressure is increased to 43 GPa.

It is well known that in Ce systems exhibiting Kondo effect phenomena, the
Kondo temperature $T_{\text{K}}$ increases rapidly with pressure
\cite{schilling1}. In the dilute magnetic alloy (La$_{0.994}$Ce$_{0.006}%
$)B$_{6}$ the electrical resistivity decreases with increasing temperature
from its giant unitarity limit value and passes through a minimum near
$T_{\min}\approx$ 20 K before rising rapidly as the phonons of the stiff
LaB$_{6}$ host lattice become thermally excited; the Kondo temperature in this
dilute magnetic alloy is estimated to lie near 1 K \cite{winzer1}. The
increase in $T_{\text{K}}$ with pressure comes from the increasing
hybridization between the 4\textit{f} and conduction electrons that enhances
the negative covalent mixing exchange interaction responsible for the Kondo
phenomena. As $T_{\text{K}}$ increases with pressure, the temperature of the
resistivity minimum $T_{\min}$ in a dilute Kondo system would be expected to
first slowly increase with pressure, but then begin to decrease after
$T_{\text{K}}$ becomes larger than $T_{\min}$ \cite{schilling1}$.$

 \begin{figure}[b]
\includegraphics[width = 8.5 cm]{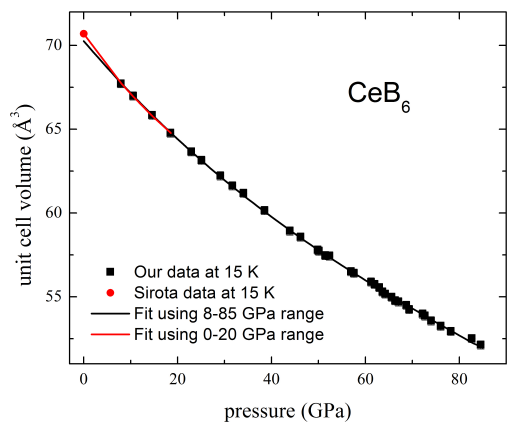} \caption{\label{fig4}(color online) Pressure dependence of unit cell
volume of CeB$_{6}$ at 15 K. Error bars are smaller than symbols. Data are fit
3rd order Birch-Murnaghan equation of state. Black curve fits present data
8-85 GPa. Red curve fits 4 data points 8-20 GPa plus point at 0 GPa from
Sirota \textit{et al.} in Ref \cite{7G}. Fit parameters are given in Table I.}
\end{figure}

\begin{table}[t]
\caption{\label{param} Summary of literature values of unit cell volume of CeB$_{6}$ at ambient pressure $V_{o},$ bulk modulus $B_{o},$ and pressure derivative of bulk modulus $B_{o}^{\prime}$ compared to present results (see
Fig.~4).}
\begin{ruledtabular}
\begin{tabular}{c c c c c c c}
Method & $P$(GPa) & Reference & $T($K$)$ & $V_{o}($\AA $^{3})$ & $B_{o}%
$(GPa) & $B_{o}^{\prime}$ \\ \hline
x-ray & 8-85 & this paper & 15 & 70.3(3) & 204(10) & 2.5\\
x-ray & 0-20 & this paper & 15 & 70.69(9) & 167(8) & 5.4\\
x-ray & 0 & \onlinecite{7G} & 15 & 70.7 & - & -\\
x-ray & 0-20 & \onlinecite{8G} & 300 & 71.27 & 166 & 3.2\\
x-ray & 0-10 & \onlinecite{9G} & 300 & 70.86 & 159 & -\\
ultrasound & 0 & \onlinecite{10G} & 10 & - & 191 & -\\
ultrasound & 0 & \onlinecite{11G} & 300 & - & 168 & -\\
Bril. scatt. & 0 & \onlinecite{11G} & 300 & - & 182 & -\\
DFT & 0 & \onlinecite{12G} & 300 & 72.47 & 173 & 3.9\\
DFT & 0 & \onlinecite{13G} & 0 & 71.16 & 162 & -\\
\end{tabular}
\end{ruledtabular}
\end{table}

In the dense Kondo compound CeB$_{6},\ T_{\min}$ lies near $110$ K at both
ambient \cite{Slu,mori1} and 0.5 GPa pressure (see Fig. 2). The higher value
of $T_{\min}$ for CeB$_{6}$ (110 K) compared to that for the dilute magnetic
alloy (20 K) is due to the much higher Ce concentration in the former. The
marked resistivity maximum in CeB$_{6}$ arises from magnetic Ce-Ce
interactions and coherence effects in the Kondo lattice. Due to the increase
of its Kondo temperature with pressure, $T_{\min}$ for CeB$_{6}$ increases
with pressure from 0.5 GPa to 7.7 GPa to 12 GPa, disappearing completely at 20
GPa and above. However, it is difficult to understand why the resistivity of
CeB$_{6}$ does not increase over the measured temperature range as the Kondo
temperature moves to ever higher temperatures, as would be expected for a
dilute magnetic Kondo alloy \cite{schilling1}$.$

The unexpected overall rapid decrease in $R(T)$ for CeB$_{6}$ as the pressure
increases to 43 GPa could be taken to suggest that a valence transition occurs
in the Ce cation. A pressure-induced change of Ce's valence from Ce$^{+3}$ to
Ce$^{+4}$ would leave the Ce cation devoid of 4$f$ electrons and, therefore,
completely quench the dense Kondo state, whereby $R(T)$\ would decrease
significantly. At 43 GPa the temperature dependence of the resistivity looks
much like that of a conventional Fermi liquid metal. Unfortunately, the
signal/noise ratio in this experiment is not sufficient to establish whether
or not the relation $\rho(T)\propto T^{2}$ holds, as one would expect from a
Fermi liquid at sufficiently low temperatures.

The valence-change scenario receives some support from the fact that the
temperature-dependent part of the resistivity at 43 GPa can be estimated
\cite{estimate1} to change by only $\Delta R\approx$ 9 $\mu\Omega$cm from 295
K to 1.3 K. This value of $\Delta R$ is well below than that found for pure La
\cite{legvold}, which contains no 4$f$ electrons, and is only about six times
greater than that for pure copper, one of the best conductors known. It would
be of considerable interest to carry out XANES measurements on CeB$_{6}$ to
pressures of at least 50 GPa to establish whether or not Ce in CeB$_{6}$
undergoes a valence change from Ce$^{+3}$ to Ce$^{+4}$.

In summary, electrical resistivity measurements on CeB$_{6}$\ from 1.3 K to
295 K to pressures as high as 122 GPa fail to find any evidence for a
transition into a topological insulating state. However, pressures of 43 GPa
are found to completely transform CeB$_{6}$ from a dense Kondo system into
what appears to be an ordinary Fermi liquid metal. Future high-pressure
synchrotron spectroscopy studies are recommended to shed light on the exact
nature of this transformation and to establish whether or not it is driven by
an increase in valence.\vspace{0.53cm}

\begin{acknowledgments}
The authors would like to express gratitude
to T. Matsuoka and K. Shimizu for sharing information on their high-pressure
electrical resistivity techniques used in this study. Thanks are due A.
Gangopadhyay for his critical reading of the manuscript. This work was
supported by the National Science Foundation (NSF) through Grant No.
DMR-1104742 and by the Carnegie/DOE Alliance Center (CDAC) through NNSA/DOE
Grant No. DE-FC52-08NA28554. Work at Argonne National Laboratory is supported
by the U.S.Department of Energy, Office of Science, under contract No. DE-AC02-06CH11357.
\end{acknowledgments}


\begin{thebibliography}{99}
\bibitem {estimate1}From Ref \cite{mori1} the value of the resistivity at
$T_{\min}$ at ambient pressure is $\sim$78 $\mu\Omega$cm, thus allowing us to
calibrate the resistance values in Fig. 2.
\end{thebibliography}
\end{document}